\documentclass[sigconf,nmanuscriptatbib=true,anonymous=false,nonacm]{acmart}

\usepackage{amsmath,amssymb,amsfonts}
\usepackage{graphicx}
\usepackage{textcomp}
\def\BibTeX{{\rm B\kern-.05em{\sc i\kern-.025em b}\kern-.08em
    T\kern-.1667em\lower.7ex\hbox{E}\kern-.125emX}}

\usepackage{url}
\usepackage{xspace}
\usepackage{multirow}
\usepackage{subfigure}
\usepackage{soul}
\usepackage{adjustbox}
\usepackage{enumitem}
\usepackage{adjustbox}

\usepackage{amsthm}%
\usepackage{mathrsfs}%
\usepackage{graphicx}%
\usepackage{multirow}%

\usepackage{xcolor}%
\usepackage{textcomp}%
\usepackage{manyfoot}%
\usepackage{booktabs}%
\usepackage{algpseudocode}%
\usepackage{listings}%

\newcommand{\knn}{\ensuremath{k}NN\xspace}
\newcommand{\componentQueueNode}{\textit{queue-node}\xspace}
\newcommand{\componentCpPartialDistance}{\textit{partial-distance}\xspace}
\newcommand{\componentCpVectorAdder}{\textit{vector-adder}\xspace}
\newcommand{\componentCpAdder}{\textit{full-adder}\xspace}
\newcommand{\componentCp}{\textit{distance-computation}\xspace}

\newcommand{\cast}{\textsc{CAsT}\xspace}

\newcommand{\seq}{\texttt{SequentialQ-CPU}\xspace}
\newcommand{\bcpu}{\texttt{BatchQ-CPU}\xspace}
\newcommand{\sqcpu}{\texttt{SingleQ-CPU}\xspace}

\newcommand{\gist}{\textsc{GIST}\xspace}
\newcommand{\yfcc}{\textsc{YFCC100M-HNfc6}\xspace}
\newcommand{\treccast}{\textsc{MS-MARCO}\xspace}

\usepackage{soul}
\usepackage{xcolor}
\usepackage{xcolor}

\def\textbf#1{{\bfseries #1}}

\begin{document}


\title{Exact Nearest-Neighbor Search on Energy-Efficient FPGA Devices}

\author{Patrizio Dazzi}
\affiliation{%
    \institution{University of Pisa}
    \department{National Research Council of Italy}
   \city{Pisa}
   \country{Italy}
}

\author{William Guglielmo}
\affiliation{%
    \institution{ISTI-CNR, University of Pisa}
    \department{National Research Council of Italy}
   \city{Pisa}
   \country{Italy}
}

\author{Franco M. Nardini}
\affiliation{%
    \institution{ISTI-CNR, Pisa}
    \department{National Research Council of Italy}
   \city{Pisa}
   \country{Italy}
}

\author{Raffaele Perego}
\affiliation{%
    \institution{ISTI-CNR, Pisa}
    \department{National Research Council of Italy}
   \city{Pisa}
   \country{Italy}
}

\author{Salvatore Trani}
\affiliation{%
    \institution{ISTI-CNR, Pisa}
    \department{National Research Council of Italy}
   \city{Pisa}
   \country{Italy}
}
\orcid{0000-0003-0194-361X}
\renewcommand{\shortauthors}{}




\begin{abstract}
This paper investigates the usage of FPGA devices for energy-efficient exact \knn search in high-dimension latent spaces. This work intercepts a relevant trend that tries to support the increasing popularity of learned representations based on neural encoder models by making their large-scale adoption greener and more inclusive.
The paper proposes two different energy-efficient solutions adopting the same FPGA low-level configuration.
The first solution maximizes system throughput by processing the queries of a batch in parallel over a streamed dataset not fitting into the FPGA memory.
The second minimizes latency by processing each \knn incoming query in parallel over an in-memory dataset. 
Reproducible experiments on publicly available image and text datasets show that our solution outperforms state-of-the-art CPU-based competitors regarding throughput, latency, and energy consumption.
Specifically, experiments show that the proposed FPGA solutions achieve the best throughput in terms of queries per second and the best-observed latency with scale-up factors of up to 16.6$\times$. 
Similar considerations can be made regarding energy efficiency, where results show that our solutions can achieve up to 11.9$\times$ energy saving w.r.t. strong CPU-based competitors.
\end{abstract}



\maketitle

\section{Introduction}
\label{sec:intro}
Top-$k$ Nearest Neighbor (\knn) search in multi-dimension spaces is a well-known problem, extensively studied in the literature. 
Given a  query object $q \in \mathbb{R}^{d}$, a collection of objects of the same dimensionality and a distance function, it asks for finding the $k$ objects in the collection having the smallest distance to the query. \knn search is a fundamental primitive in several areas, such as databases, information retrieval, recommender systems, pattern recognition, machine learning, and statistical data analysis. Despite the vast amount of studies discussing optimized algorithms and indexing data structures for \knn search, the problem of efficiently finding the data points closest to a given query is still a hot research topic. Its importance is due to the large number of applications and the growing importance and availability of complex data represented in high-dimension spaces. 
Most of the efforts to make \knn search efficient and scalable address the problem by relaxing it and searching for data points that are likely to be the nearest neighbor ones, i.e., by accepting some loss in accuracy with Approximate NN (ANN) search methods~\cite{TKDE2020}. 
Indexing data structures for fast ANN search rely on quantization or hashing/binning techniques~\cite{pan2020product,andoni2008near}. Such techniques typically cluster the data points in the collection based on the distance function and perform a search in two steps: first, the centroids of the clusters closest to the query are identified; second, a subset of clusters associated with the nearest centroids is visited exhaustively to determine the closest data points. 
Due to the curse of dimensionality, visiting all neighboring clusters is impractical, but any heuristics limiting the number of neighboring clusters visited may even severely impact the recall of these approaches~\cite{TKDE2020}. 
Some recently proposed graph-based methods achieve competitive recall rates for ANN search~\cite{GGNN,NEURIPS2020}. They work by building a partial representation of the \knn graph linking the points in the search space to their nearby points. Retrieval starts from a random node in the graph and proceeds iteratively by following at each step the edge connecting the node closest to the query, if any. 

All ANN search methods have to trade-off between accuracy and computational efficiency. On the other hand, exact \knn search over billions of high-dimension data points is challenging even when leveraging massive GPU parallelism~\cite{TKDE2020} and sophisticated metric indexes~\cite{exactACMCS}.
However, many applications require exact \knn search on collections of manageable size. These include, for example, public safety applications~\cite{exactACMCS}, data cleaning~\cite{VLDB06}, near-duplicate detection~\cite{NDD12}, and bioinformatics \cite{SIGMOD11}. 
Given the high construction costs and the limited efficiency gain of metric indexes on small/medium datasets, a competitive method for retrieving \knn results is to perform an exhaustive search, i.e., comparing the query object with all the points in the collection with a brute force approach. Moreover, even when datasets are huge,  either \knn or ANN search methods based on hashing/binning techniques rely on an exhaustive search to visit selected partitions of the search space. 

In this paper, we investigate using FPGA devices for energy-efficient \knn search. We propose two flexible solutions based on the same FPGA configuration, and we showcase their efficiency on high-dimension real-world datasets for image and text retrieval. Our work intercepts a relevant trend in the scientific community aimed at making query processing greener \cite{10.1145/3477495.3531766} and AI models more efficient, sustainable and inclusive~\cite{grenAI,parrots}.   
 
%
In summary, we focus on sustainable AI inference using resource-constrained, energy-efficient devices and we contribute with:
\begin{itemize}
    \item two different FPGA solutions for energy-efficient AI inference based on exact \knn search in high-dimensional spaces. The first solution maximizes system throughput by processing in parallel a batch of \knn queries over a streamed dataset not fitting into the FPGA memory. The second one minimizes latency by processing a stream of incoming \knn queries over an in-memory dataset. 
    \item a rigorous methodology for assessing the proposed solutions with reproducible experiments conducted on publicly available image and text retrieval datasets with vectors varying from $769$ to $4,960$ dimensions. Our proposal's throughput, latency, and energy consumption are compared with those of the optimized multi-threaded implementation available in the FAISS \knn library~\cite{faiss}. Results show that our FPGA solutions achieve the best throughput and latency, while enabling up to $11.9\times$ energy saving compared to a high-end multi-core CPU. Moreover, we show that the FPGA solution remarkably outperforms even a high-end GPU in application scenarios where it is not possible to batch the queries in large sets to fully exploit the GPU cores.
\end{itemize}

The rest of the paper is organized as follows.
Section~\ref{sec:related} discusses the related work while
Section~\ref{sec:knn} introduces our FPGA solution and details the logic organizations of its components.
Section~\ref{sec:experiments} details the experimental settings and discusses the results of the experiments conducted.
Finally, Section~\ref{sec:conclusions} concludes the work and draws some future lines of investigation.


\section{Related Work}
\label{sec:related}
We restrict the survey of related work to the areas that are more relevant to our work, specifically the approaches for accelerating exhaustive NN search using parallel computational platforms, e.g., multi-threaded CPU~\cite{7073244}, GPUs~\cite{4563100,faiss}, and FPGA devices. Lu~\emph{et al.} design and implement CHIP-KNN, a high-performance \knn accelerator, which optimizes the off-chip memory access on cloud FPGAs with multiple High-Bandwidth Memory (HBM) banks~\cite{9415564}. Given a user configuration of the \knn parameters, CHIP-KNN automatically generates the design of the optimal accelerator on a specific FPGA platform. Experiments on the Xilinx Alveo U200 FPGA show that, compared to a $16$-thread CPU implementation, CHIP-KNN achieves up to a $19.8\times$ performance speedup with up to $16.0\times$ performance/cost improvements.
 
In the same line, Bank Tavakoli~\emph{et al.} propose RPkNN, a framework leveraging OpenCL to accelerate nearest neighbors (NN) search on FPGA devices~\cite{9714070}. The authors do that by complementing the approach on FPGA with random projection dimensionality reduction. RPkNN enables parallel \knn computations with a single memory channel and takes advantage of the sparsity features of the input data to implement a highly optimized and parallel implementation of random projection. The authors compare RPkNN implemented on the Samsung SmartSSD CSD with the kNN implementation of the scikit-learn library running on an Intel Xeon Gold 6154 CPU. The experimental results show that the proposed RPkNN solution achieves a latency speedup of $46\times$ on a single kNN computation. Moreover, experiments show that RPkNN is $1.7\times$ faster than a previous state-of-the-art FPGA-based method.

Vieira~\emph{et al.} propose two adaptive FPGA architectures of the \knn classifier~\cite{8911384}. The two architectures employ a novel dynamic partial reconfiguration (DPR) architecture of the \knn, which allows for an efficient dynamic implementation on FPGA for different \emph{k}'s. Experiments show that kNN-STUFF with $24$ accelerators achieves performance improvements up to $67.4\times$ compared to an optimized software implementation of the \knn running on a single core of the ARM Cortex-A9 CPU. Furthermore, kNN-STUFF achieves an energy reduction of up to $50.6\times$.
Such results are not completely surprising as FPGAs are more performing and energy-efficient than CPUs and GPUs in different domains. For instance, Li~\textit{et al.}\cite{10.1145/3370748.3406567} presents an acceleration framework, FTRANS, for transformer-based large-scale language representations, targeting Natural Language Processing, which makes use of FPGAs. 
Other applications in Artificial Intelligence have a clear benefit in terms of energy efficiency due to the exploitation of FPGA-based solutions. Some of those are particularly suitable for FPGAs due to their vocation to be performed on embedded devices. SparkNOC~\cite{XIA2021101991}, standing for SparkNet on Chip, is an FPGA-based accelerator implemented on an Intel Arria 10 that maps all the layers of lightweight neural network architecture on the FPGA. SparkNOC outperforms the original approach in terms of energy consumption: $44.48$ GOP/s/W in place of the $0.086$ provided by Intel Xeon E5 and $0.157$ of a Titan~X GPU. As pointed out by Podobas \textit{et al.}~\cite{10.1145/3468044.3468052}, the flexibility that FPGAs have in terms of number formats and data representation leads such devices to be more energy efficient than CPUs and GPUs, even when compared to more recent devices. 
A similar trend can be observed in embedded systems for computer vision. Qasaimeh \emph{et al.}~\cite{8782524} compare the energy efficiency of CPU, GPU and FPGAs in kernels of computer vision applications (implemented in OpenCV, VisionWorks and xfOpenCV, respectively). The comparison results clearly show that FPGA outperforms the other platforms in terms of energy efficiency in most experiments, measured on a per-frame granularity.

Beyond being exploited in resource-constrained devices, FPGAs are becoming an important component of high-performance computing platforms~\cite{9556330}, which are becoming increasingly heterogeneous to fulfill the different needs of modern HPC applications. The involvement of such devices is key for some application types. 
The generalized application of FPGAs in place of CPUs and GPUs for HPC applications has been demystified by recent studies~\cite{10.1002/cpe.6570}. 
These studies have illustrated the considerable advantages of leveraging FPGAs, particularly with respect to energy efficiency, in targeted High-Performance Computing (HPC) scenarios. Specifically, FPGAs prove highly beneficial in memory-intensive streaming computations characterized by predetermined spatial and temporal locality patterns. They excel in scenarios involving pipelined computations that without massive fine-grained data parallelism, as well as in tasks with stringent latency requirements and those demanding network-intensive computations.

\paragraph*{Our Contributions}
The work presented in this paper is a contribution toward energy-efficient exact \knn on FPGA in high-dimensional spaces. We propose two novel FPGA architectures. While the first configuration maximizes system throughput by processing batches of \knn queries, the second solution minimizes latency by processing, once at a time, a stream of incoming \knn queries over an in-memory dataset.
The work closest to ours is CHIP-KNN \cite{9415564}, where an exact FPGA \knn architecture is proposed. Unlike our work, CHIP-KNN  does not consider the streaming vs. in-memory scenarios and only addresses low-dimensional spaces. On the other hand, we propose two extremely optimized solutions for throughput or latency. These solutions perfectly fit different information retrieval use cases -- e.g., dense retrieval with high-dimensional learned representations and content-based image retrieval -- where we often have to perform both high-throughput and low-latency online query processing tasks. 

\section{Designing \knn on FPGA devices}
\label{sec:knn}

We aim to efficiently answer \knn queries in high-dimensional spaces by exploiting an FPGA device to minimize energy consumption. Let $\delta$ be a {\em distance function}, $\delta: \mathbb{R}^{d} \times \mathbb{R}^{d} \to \mathbb{R}$,  measuring the distance between two vectors in $\mathbb{R}^{d}$. Given a query  $q \in \mathbb R^d$ and a dataset  $\mathcal{D} = \{x_1,x_2,\ldots,x_n\}, x_i \in \mathbb{R}^{d}$, we  want to  retrieve the $k$ vectors in $\mathcal{D}$ that are the closest to $q$ according to the distance function $\delta$, i.e., the $k$ nearest-neighbor vectors denoted as \knn(q).


We propose two different solutions for performing \knn search on FPGA devices. 
The first solution, called FQ-SD (Fixed Queries, Streamed Dataset), is optimized to answer a fixed batch of \knn queries over a dataset that does not fit in the memory of the FPGA device. The dataset is, thus, split into disjoint partitions fitting in the FPGA memory that are streamed, one at a time, to the device from the host. 
FQ-SD maximizes system throughput by processing the whole batch of queries in parallel over the input stream.

The second logical configuration, FD-SQ (Fixed Dataset, Streamed Queries), targets those scenarios in which the dataset entirely fits in the memory of the FPGA device. FD-SQ is thus optimized to speed up processing a stream of incoming \knn queries over an in-memory dataset. 

Since the \knn search process implemented is exact and exhaustive, both solutions do not introduce particular constraints on the distance function, e.g., $\delta$ can be a metric distance or not. Moreover, it is worth highlighting that FQ-SD and FD-SQ are logical configurations based on the same FPGA hardware configuration. The actual behavior of the FPGA architecture configuration is selected at run-time by activating the same kernels with different sets of parameters establishing the way data flows into the components. Sharing the same hardware configuration allows the host to activate FQ-SD and FD-SQ solutions interchangeably without having to pay the overhead of flashing the FPGA hardware for changing its behavior. 

In the following, we introduce the FPGA technology and the two logical configurations optimized for \knn throughput and latency, respectively. Then, we detail the functional building blocks for \knn, emphasizing how they are optimized for efficient hardware synthesis. 

\subsection{Using Programmable Logic}
The features of current FPGA devices allow their adoption for \sloppy high-performance computing tasks such as \knn search. 
A modern FPGA device integrates on a single chip a high-capacity Programmable Logic unit, including banks of memories, control, and logic components like Flip Flops or Look-up Tables. These components are used to implement Intellectual Property blocks that execute the algorithms on the FPGA. The amount of components available in the FPGA is limited, and this limitation, in turn, constrains the design and the deployment of the algorithm, which usually involves a trade-off between processing speed and resource utilization.
To fully leverage FPGA logic, data processing should occur in parallel, minimizing reliance on jumps and recursive calls.
To this end, High-Level Synthesis (HLS) tools are used to create hardware from a high-level description, using directives to specify concurrency and pipelining opportunities.
The HLS tool translates the code to a Register Transfer Level specification of the hardware and returns an estimation of execution latency and resource utilization. In this way, the designer can broadly evaluate the performance of different implementation strategies before deploying them on the hardware. In this phase, the designer is also asked to detail the data communication between FPGA and the host computer.

\subsection{Logical configurations for FPGA \knn}
\label{sec:logicalconf}

\begin{figure}
\centering
\includegraphics[width=1.0\linewidth]{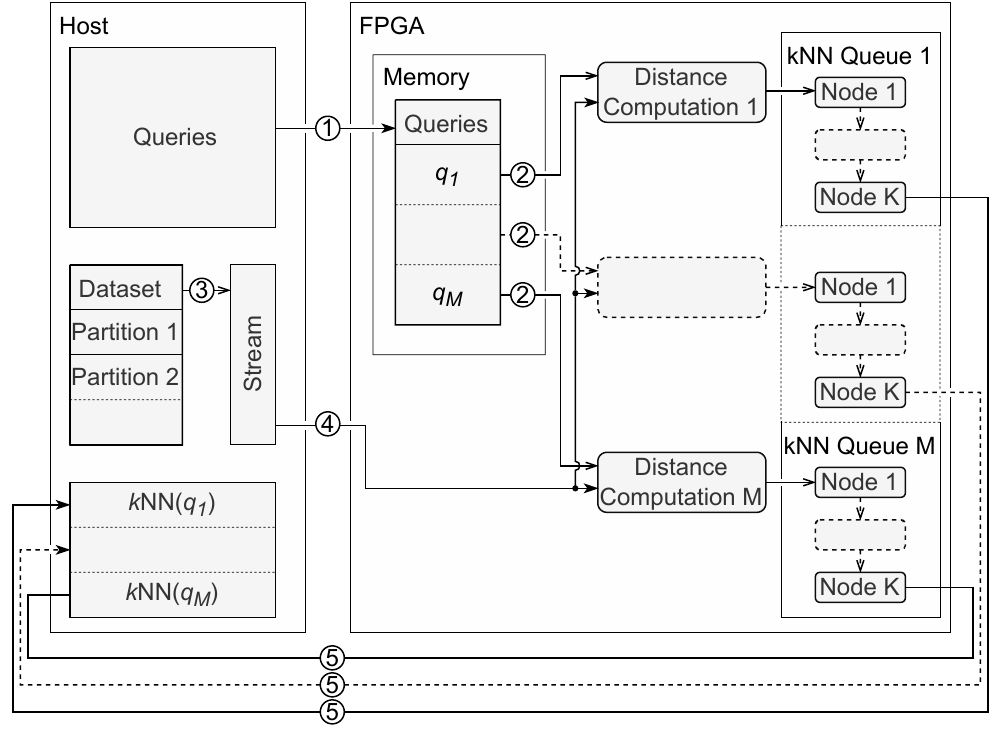}
\caption{The logical architecture of the FQ-SD configuration.\label{fig:FQ-SD}}
\end{figure}

\begin{figure}[htb]
\centering
\includegraphics[width=1.0\linewidth]{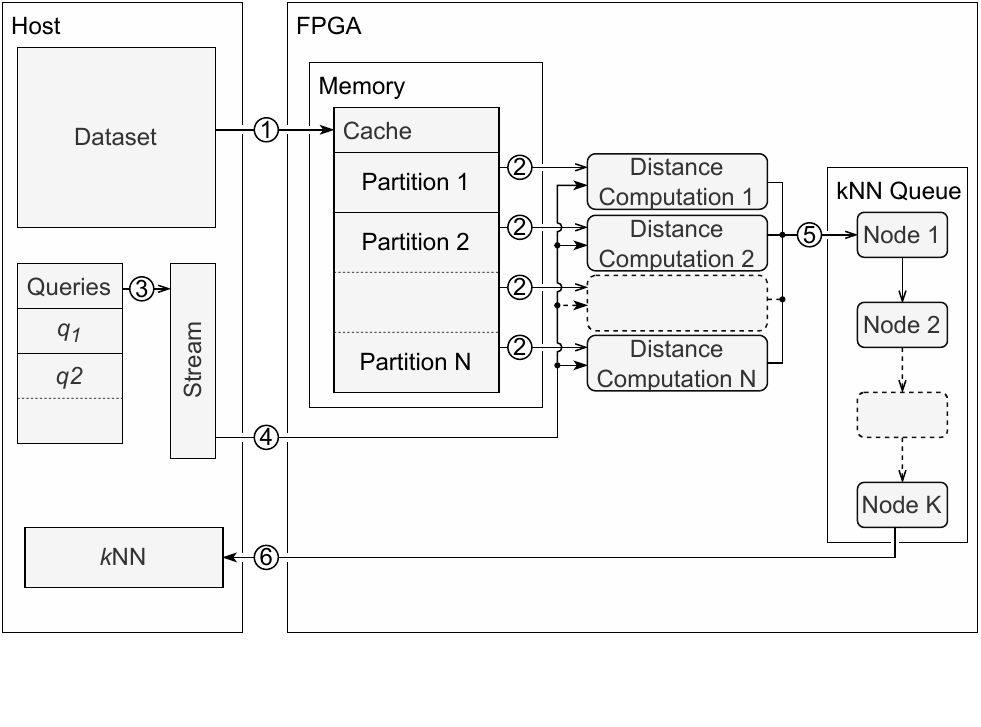}
\caption{The logical architecture of the FD-SQ configuration.\label{fig:FD-SQ}}
\end{figure}

Our FQ-SD and FD-SQ solutions, optimized for \knn throughput and latency, are depicted in Figure~\ref{fig:FQ-SD} and Figure~\ref{fig:FD-SQ}, respectively. Numbered arrows depict the flows of data across logical components.

\paragraph*{FQ-SD: Fixed Queries, Streamed Dataset}
The FQ-SD solution aims at answering a fixed batch of $M$ \knn queries, efficiently, over a dataset that does not fit in the FPGA memory. 
The $M$ queries are first loaded into the FPGA memory (arrow 1). The $M$ distance computation instances shown in the figure read the $M$ query vectors from the FPGA memory, one query for each instance (arrow 2). The dataset resides on the host computer that splits it into $N$ disjoint partitions of equal size fitting in the FPGA memory and aligned to the FPGA data transfer width with padding when needed. The host streams the $N$ partitions to the FPGA (arrows 3 and 4) that processes one partition at a time by orchestrating the computation of the distances between the $M$ queries and all the vectors in each partition. Each distance between a vector $x_j$ and the query $q_i$, i.e., $\delta(x_j,q_i)$, is stored into the $i$-th \knn queue.
Each \knn queue implements a $\min$ heap storing the indices and the distances of the $k$ closest vectors in $\mathcal{D}$, i.e., the set \knn($q_i$). Finally, the \knn results for all the queries are transferred to the host (arrow 5).
It should be noted that the hardware configuration contains just one $min$ heap storing $k$ indices, implemented with a pipeline of $k$ elements, one element per each stored index.
To enable the computation of a batch of queries, we need to keep track of the nearest neighbors for each query individually.
For this reason, we added the possibility of logically partitioning the pipeline of $k$ elements into $M$ pipelines of $k/M$ elements, each implementing an independent min heap storing $k/M$ indices.
This way, executing the FQ-SD solution for a single query looking for the $k$ nearest neighbors or a batch of $M$ queries looking for the $k/M$ nearest neighbors in parallel is possible. The two versions are implemented with the same hardware configuration, and the host can interchangeably choose which of the two to use at run-time without the need to flash the FPGA.


\paragraph*{FD-SQ: Fixed Dataset, Streamed Queries}

The design of the FD-SQ solution covers the case in which the dataset of multi-dimensional vectors entirely fits into the memory available on the FPGA device. The dataset is first split into $N$ partitions completely loaded into the FPGA memory (arrow 1). Also in this case, the data in each partition is padded if the size of an FPGA memory block is not exactly a multiple of the size of the multi-dimensional vectors. The \knn queries are streamed, one at a time, from the host to the FPGA device (arrows 3 and 4). The distances between the current query $q_i$ and the vectors of the memory-resident dataset are computed in parallel by $N$ instances of the distance computation component, one for each partition. This component iterates by: i) reading from the respective partition stored in the FPGA memory one vector $x_j$  (arrow 2); ii) computing the distance $\delta(x_j,q_i)$; iii) inserting $\delta(x_j,q_i)$ and the reference to $x_j$ into a common \knn queue selecting the top-$k$ nearest-neighbor vectors. Finally, the results in \knn($q_i$) are transferred to the host (arrow 5), and the computation of the next query $q_{i+1}$ in the incoming stream starts.


\subsection{Building blocks for efficient FPGA \knn}
\label{sec:knnfpga}
We detail the design of three main FPGA building blocks implementing the proposed \knn solutions: \textit{double buffering}, \componentCp, and \knn \textit{queue}. While double buffering is used to optimize dataset streaming in the FQ-SD solution, the other two building blocks are FPGA components used both in FQ-SD and FD-SQ logical configurations. 

\paragraph*{Double buffering for dataset streaming}
Double buffering is a technique aimed at avoiding read/write conflicts.
A read/write-access conflict can occur in the FQ-SD solution when: i) the host is still transferring a partition of data when the FPGA accesses it to compute distances; ii) the FPGA is still reading the vectors from a partition when the host wants to overwrite the transfer buffer.
Such access conflicts impact on the data transfer bandwidth by causing the host or the FPGA to be idle, waiting for one another.
For the two devices to work efficiently, in parallel, the host should work at least one partition ahead of the FPGA by exploiting multiple dynamic buffers so that the host writes partition $n+1$ while the FPGA reads and compute distances for the data in partition $n$.
We exploit a double buffering technique to maximize the data transfer bandwidth in the FQ-SD solution. The host stores in memory the dataset partitions that should be transferred to the FPGA in specific data structures aligned to the width of the host-FPGA transfer buffer.
A partition $i$ is transferred by the host to the FPGA at time $T_i$ by writing it in the FPGA memory bank $(i \mod 2) + 1$.
The partition transferred to the FPGA at time $T_{i}$ is read by the FPGA logic for distance computations from memory bank $(i \mod 2) + 1$ at time $T_{i+1}$.
This double buffering schema allows the bandwidth of the Host-FPGA connection to be fully exploited by preventing the host and the FPGA from idle waiting. 
Specifically, the data is ingested into an FPGA device connected via a PCI-e bus ($16\times$) at $12.5$ GB/s, a speed quite close to the maximum theoretical bandwidth of $16$ GB/s.



\paragraph*{Distance computation}
The \componentCp component
computes the distance between a query $q$ and a vector $x$ in $\mathbb{R}^{d}$. In detail, \componentCp consists of three sub-components: \componentCpPartialDistance, \componentCpVectorAdder, and \componentCpAdder, each one implemented in a separate pipeline to limit resource usage. In the following, we refer to the squared euclidean distance, but different distances can be computed by slightly changing the implementation of these components.
The \componentCpPartialDistance subcomponent fetches the query $q$ and partitions it in $r$ parts of $w$ elements, $r = \lceil\frac{d}{w}\rceil$ and $w$ is determined on the basis of the amount of data that can be read from the FPGA memory bank in a single operation.
The current dataset vector $x$ is also partitioned in $r$ parts, with the same alignments of $q$; the distance is computed for each pair of partitions $q$/$x$, and $r$ distances are totally computed for each dataset vector.
The distances computed by this component are stored in an array $A$ of $m$ shift registers; once $m$ distances are computed, $A$ is copied and sent to \componentCpVectorAdder.
The value of $m$ depends on both the FPGA hardware and the compiler version used. It must be chosen accordingly to the output (the ``Initiation Interval'') produced by the compiler for the \componentCpVectorAdder pipeline with an initial value of $m = 1$. The final value of $m$ is usually small, and we obtained $m = 8$ for our environment.
The \componentCpVectorAdder component stores an array $B$ of $m$ elements initially set to zero. It performs the vector addition $B$ = $B$ + $A$, with $A$ the array received as input by the previous component. Every $r'$ activations, with $r' = \lceil\frac{r}{m}\rceil$, it has processed all the partitioned distances of $x$. Then it copies $B$, sends the copy to \componentCpAdder and sets $B$ to zero.
The \componentCpAdder subcomponent is activated once per query/vector pair. For each pair, it receives an array of $m$ elements and sums up all its values obtaining the final squared euclidean distance between the query and the current vector.
The rationale for realizing \componentCp with three pipelines, \componentCpPartialDistance activated $r$ times for each query/vector pair and \componentCpVectorAdder $r'$ times, relies on instructing the compiler to create a simpler and less resource-hungry hardware configuration for these specific functions, thus saving FPGA resources.

\paragraph*{\knn queue}
The \knn queue component is a pipeline of $k+2$ simple elements, with $k$ being the cutoff of the \knn search. 
The first element is a \textit{reader} receiving from a \componentCp component an end-of-stream marker or a pair $(\delta(x_i,q),i)$, with $\delta(x_i,q)$ being the distance between a vector $x_i$ and the current query $q$. In both cases, the data received is forwarded to the next element.
Each of the $k$ following \componentQueueNode elements stores internally a pair $(\delta_{min},i_{min})$, and they all have the same behavior: they receive from the previous element an end-of-stream marker or a pair $(\delta_{new},i_{new})$ that can be marked as a solution, i.e., a pair belonging to the \knn set or not.
In the former case, the stored pair is marked as a solution and sent to the next element, followed by an end-of-stream marker.
In the latter case, the element can perform two different operations:
\begin{enumerate}[label=(\Alph*)]
    \item \label{enumerate:queue-operations:store} it sends the pair $(\delta_{min},i_{min})$ to the next element and stores $(\delta_{new},i_{new})$ as the current pair;
    \item \label{enumerate:queue-operations:forward} it sends the pair $(\delta_{new},i_{new})$ to next element.
\end{enumerate}
If the pair is not marked as a solution, the value of $\delta_{new}$ is compared with the value $\delta_{min}$ currently stored in the element: it performs the~\ref{enumerate:queue-operations:store} operation if $\delta_{new} < \delta_{min}$, otherwise it performs the \ref{enumerate:queue-operations:forward} operation.
If the new pair is marked as a solution, the element marks its stored pair as a solution and applies the \ref{enumerate:queue-operations:store} operation without any comparison.

The last element of the pipeline is a \textit{writer} receiving pairs from the last \componentQueueNode element. All the pairs not marked as a solution that \textit{writer} receives are dropped, while the ones marked as belonging to the \knn set are stored in an array on the FPGA memory in the reversed order they are received.

We get that until the \textit{reader} sends an end-of-stream marker, each \componentQueueNode keeps track of the pair containing the minimum distance it has received and does not forward it.
As the \textit{reader} emits the end-of-stream marker, the solutions stored by the \componentQueueNode elements start to flow toward the \textit{writer}.
The termination process of the \componentQueueNode elements involves two phases:
\begin{enumerate}
    \item The \componentQueueNode receives  a solution pair: it marks its current pair as a solution,  forwards it to the next node, and stores the received solution as its new pair;
    \item The \componentQueueNode receives the end-of-stream marker: it marks its current pair as a solution, forward it to the next node, and terminates.
\end{enumerate}
Note that the first node only executes the second phase, and all the nodes execute the second phase only once.
The first phase is instead executed one time by the second node, two times by the third node, three times by the fourth node, and so on.
\section{Experimental Evaluation}
\label{sec:experiments}
This section details the experimental evaluation to compare our proposals against state-of-the-art competitors. Experiments are designed to answer three main research questions:

\begin{itemize}
\item \textbf{RQ1}: What is the computational efficiency of our FQ-SD and FD-SQ solutions on different high-dimension datasets compared to FPGA competitors and CPU baselines?
\item \textbf{RQ2}: How much energy do our FPGA solutions save compared to CPU implementations having similar or lower performance?
\item \textbf{RQ3}: Given a fixed amount of FPGA programmable logic, can we further improve the performance by trading off between the degree of parallelism and cutoff $k$ of the \knn search? How does it perform in comparison with GPU-based \knn solutions?
\end{itemize}

In the following, we introduce i) the two use cases selected to showcase our FPGA \knn solution, ii) the public datasets employed in the evaluation, iii) the experimental infrastructure used, and iv) the state-of-the-art baselines used for comparison.
Finally, we discuss the comprehensive set of experiments conducted to answer the above research questions.

\subsection{Use Cases}
\label{sec:usecases}
This section details the two use cases chosen to assess our energy-efficient solution for FPGA \knn search.

\paragraph*{Content-based image retrieval} The first use case grounds in content-based image retrieval. State-of-the-art image retrieval systems index billions of images by providing low-latency search and access services~\cite{Noh_2017_ICCV}. In this context, an important aspect becomes the design of energy-efficient and scalable platforms exploiting parallel and distributed \knn search solutions. 
Depending on the size of the image collection, the dimensionality of the image representation, and the capacity of the FPGA device used, this scenario can fit either the FD-SQ or the FQ-SD FPGA configurations.  

\paragraph*{Passage retrieval for question answering with dense retrieval models} 
Our second use case capitalizes on recent advances in Dense Retrieval (DR) models able to effectively capture the deep semantic relationship between the text of queries and documents~\cite{ance,star,dpr,10.1145/3394486.3403305,CastDR}. 
Specifically, we focus on the efficiency of \textit{passage retrieval} in question answering where, given a query, the task asks for retrieving textual passages relevant to the query. This task has also proven essential for the emerging paradigm of conversational systems, where fast query answering is required to engage the users in a fluid conversation~\cite{cachingClient-side}. Consistently with state-of-the-art proposals in this line~\cite{convdr, zamani2022conversational}, we adopt a single-representation DR model, where each document of the collection is encoded with a single---learned---vector in a multi-dimensional latent space, i.e., also known as embedding. Given a query embedded in the same multi-dimensional space, online retrieval and ranking is performed through a \knn search based on standard metrics such as maximum inner product~\cite{ance,dpr,star, sbert} or minimum Euclidean norm~\cite{colbert}.
The state-of-the-art single-representation models proposed in the literature are DPR~\cite{dpr}, ANCE~\cite{ance}, and STAR~\cite{star}. The main difference among these models is how the fine-tuning of the underlying pre-trained language model, i.e., BERT, is carried out. We selected the embeddings computed by the STAR model for our experiments since it employs hard negative sampling during fine-tuning, obtaining better representations of effectiveness w.r.t. ANCE and DPR.
STAR~\cite{star} encodes queries and documents with embeddings of $769$ dimensions\footnote{STAR encoding uses $768$ values, but we added one dimension to each embedding by applying the transformation proposed by Bachrach~\emph{et al.}~\cite{xbox} as well as by Neyshabur and Srebro~\cite{10.5555/3045118.3045323}, to exploit euclidean norms.}.
 



\subsection{Datasets}
To favor the repeatability of our experiments, we use the following publicly available datasets: \gist~\cite{jegou2010product}, \yfcc~\cite{10.1145/2983554.2983557}, and \treccast\footnote{\url{https://microsoft.github.io/msmarco/}}.

\gist is a corpus of descriptors derived from the first $100,000$ images extracted from the dataset of Torralba~\emph{et al.}~\cite{torralba200880}, while query descriptors are extracted from the Holidays image queries~\cite{jegou2008hamming}.

\yfcc~\cite{10.1145/2983554.2983557} is a corpus of deep features extracted from the \textsc{YFCC100M} dataset. \textsc{YFCC100M} has been released in 2014 as part of the Yahoo Webscope program\footnote{\url{https://webscope.sandbox.yahoo.com/}}, and it consists of approximately $99$.$2$ million photos and $0$.$8$ million videos, all uploaded to Flickr between 2004 and 2014 and published under a Creative Commons commercial or non-commercial license. The deep features contained in \yfcc are produced using the \texttt{fc6} hidden layer activation of the HybridNet deep convolutional neural network.

The MAchine Reading COmprehension v1 Passage (\treccast) is a dataset made up textual passages and the goal is to rank them based on their relevance with respect to a query. It comes with three sets of queries (train, dev-full, dev-small) and binary relevance labels reporting whether a passage provide an answer or not for the query. In this study, we employ the dev-small query set made up of $6$,$980$ queries.
Passages and queries are encoded in one single $769$-dimensional dense vector by applying STAR~\cite{star}, which generates a single embedding for each passage and query in the collections.
By doing so, we build a collection of ${\sim}8.8$M vectors used to answer the $6$,$980$ queries of dev-small set.
Table \ref{table:datasets} details the properties of the datasets in terms of the number of vectors in the collection, the number of dimensions of the representation, and the number of queries used in the evaluation.

\begin{table}
\caption{Properties of the datasets used.}
\label{table:datasets}
\centering
\begin{tabular}{crrr}
\toprule
Dataset & Vectors & Dimensions & Queries \\
\midrule
\gist & $1$,$000$,$000$ & 960 & $1$,$000$ \\
\midrule
\yfcc & $\sim100$,$000$,$000$& $4$,$096$ & $1$,$000$\\
\midrule
\treccast & $8$,$841$,$823$ & $769$ & $6,980$ \\
\bottomrule
\end{tabular}

\end{table}

\subsection{Experimental Infrastructure}
We deploy our novel architectures for exact nearest-neighbor search on a Xilinx Alveo U55C data center FPGA board\footnote{\url{https://www.xilinx.com/applications/data-center/high-performance-computing/u55c.html}}. The card exploits a PCI Express (PCIe) Gen3 x16 bus interface to communicate with the host machine, and it is equipped with 16 GiB of high-bandwidth memory (HBM2). The host machine is based on an Intel Xeon CPU E5-2683 v4 clocked at $2.10$ GHz with $16$ physical cores and $256$ GiB of RAM. The host machine runs Ubuntu Linux 20.04.4 and Vitis 2021.2.
To develop our applications, we employ the Xilinx Runtime (XRT), which allows software and hardware components to be designed using standard C/C++ annotated with directives specifying parallelism opportunities and placement choices.
The FPGA mounted on the Alveo U55C board comprises $1$,$304$ K look-up tables (LUTs), $2$,$607$ K Registers, and $9$,$024$ DSP slices.
The reported response times are obtained as an average of three different runs.

\subsection{Competitors}
We compare our solution against several state-of-the-art competitors for exact \knn search on CPU:
\begin{itemize}
\item \seq. This solution provides an optimized sequential query processing baseline, where each query is answered by computing exact \knn using a single thread running on a physical core of the CPU. 
\item \bcpu. This competitor provides multi-threaded query processing on a per-query basis, i.e., each query is answered using a thread on a physical core of the CPU. \bcpu is the natural competitor on CPU for the FQ-SD architecture as each thread is responsible for answering a specific query on a streamed dataset.
\item \sqcpu. This competitor solution exploits thread-level parallelism to speed up the processing of one query at a time. Here, the degree of parallelism controls the number of threads and dataset partitions used to perform an exact \knn search on a given query. \sqcpu is the natural competitor on CPU for the FD-SQ architecture as each thread is responsible for answering the same single query on a given partition of the dataset.
\end{itemize}

All three implementations are compiled with all optimization flags on and exploit the low-level C++ exact \knn search APIs (L2 computation and related data structures) provided by the FAISS\footnote{\url{https://github.com/facebookresearch/faiss}} library. FAISS~\cite{faiss} is the state-of-the-art implementation for efficient and scalable similarity search. Its low-level APIs for exact \knn are highly optimized and exploit the BLAS routines~\cite{9756685} as well as SSE2 and AVX2 instruction-level parallelism. We used the low-level C++ FAISS APIs to avoid possible overheads introduced by the Python interpreter that comes into play when using the standard FAISS high-level APIs.

In addition, we compare our novel architecture against a GPU-based \knn search made available in the FAISS library. Specifically, we experiment on an NVIDIA A100 GPU board with 40 GB of RAM and $6$.$912$ computing cores. The experiment uses FAISS version 1.7.3, which runs on top of CUDA 12.2.

We also compare our novel architectures against CHIP-KNN~\cite{9415564}. CHIP-KNN is a state-of-the-art proposal that automatically generates the design of an efficient FPGA accelerator, given a configuration of the \knn parameters used. The authors experiment with CHIP-KNN on a Xilinx Alveo U200 FPGA. Although the source code of CHIP-KNN has been released\footnote{\url{https://github.com/SFU-HiAccel/CHIP-KNN}}, due to software versions issues, it was not possible to compile and run it on our Xilinx Alveo U55C. These issues prevent a comprehensive comparison of CHIP-KNN against our novel architectures on the three datasets we employ. However, we resort to using the data presented by Lu~\emph{et al.}~\cite{9415564} to produce a fair comparison of our architectures against CHIP-KNN in terms of maximum throughput achieved w.r.t. the dimension of the vectors of the datasets employed and the cutoff $k$. Finally, we do not report comparisons with GPU \knn implementations because they fit mainly batch processing scenarios, and other works show they consume much more energy than FPGA \cite{energyGPU}.

\smallskip
\noindent \textbf{Reproducibility}. The source code used in our experimental evaluation, i.e., the implementation of our FPGA architectures and the CPU competitors, will be made publicly available upon publication of the work to allow the reproducibility of the results\footnote{\url{https://github.com/hpclab/knn-fpga}}.

\subsection{Evaluation Metrics}
Since this paper focuses on energy-efficient solutions for exact and exhaustive \knn search, the retrieval effectiveness is not impacted and is always maximized. For this reason, we focus our analysis on computational performance only.
In particular, we compare our FPGA solutions against state-of-the-art baselines by using three different efficiency metrics:
\begin{itemize}
\item Query latency: reported in milliseconds (msec), is the average time each technique needs to answer an exact \knn query.
\item Throughput: reported as the number of queries answered per second (Queries/sec), is the average number of \knn queries processed per second.
\item Energy efficiency: reported as the number of queries per Joule (Queries/J), is the average fraction of query processed per unit of energy consumed.
\end{itemize}
All the figures reported with the above metrics are obtained by averaging the results of three different runs for each experiment.

While the analysis against state-of-the-art baselines exploits the three efficiency metrics above, the comparison of our FPGA \knn architectures against CHIP-KNN is performed only in terms of throughput (GByte/sec). 
No direct comparison can be made regarding query latency and energy efficiency because the two experimental methodologies employ different datasets, vector sizes, and cutoff $k$. Nevertheless, the throughput of the different architectures can still be employed for a fair comparison because the boards used in the two works for the experiments are similar, i.e., both PCI-e-based Xilinx Alveo U200 and Alveo U55C FPGAs.

\subsection{Results and Discussion}
We report the experimental results of our comparisons on the three datasets employed. We compare FQ-SD and FD-SQ against \seq, \bcpu, and \sqcpu in terms of latency (msec), throughput (\# queries/sec), and energy (\# queries/Joule). 

\begin{table*}[htbp]
\centering
\caption{Experimental results on our three public datasets. We also report the scale-up factor achieved by each method w.r.t. \seq for latency, throughput, and energy efficiency.\label{table:results-rq1}}
\begin{tabular}{clcrcrcrrcrrcrr}
\toprule
\multirow{2}{*}{Dataset}
& \multirow{2}{*}{Method} && \multirow{2}{*}{Workers} && \multicolumn{1}{c}{Batch}   &&    \multicolumn{2}{c}{Latency}  &&  \multicolumn{2}{c}{Throughput}    && \multicolumn{2}{c}{Energy Efficiency} \\
&                         &&                          && \multicolumn{1}{c}{Size}    &&    \multicolumn{2}{c}{(msec/query)}   &&  \multicolumn{2}{c}{(Queries/sec)} && \multicolumn{2}{c}{(Queries/J)}       \\
\toprule
\parbox[t]{2mm}{\multirow{16}{*}{\rotatebox[origin=c]{90}{\gist}}}
& \seq                    &&                        1 &&                           1 &&           304  &             -- &&                3.3 &            -- &&                 0.075  &           -- \\
\cmidrule{2-15}
& \multirow{5}{*}{\bcpu}  &&                        2 &&                           2 &&   \textbf{334} &    0.9$\times$ &&                6.0 &   1.8$\times$ &&                 0.120  &  1.6$\times$ \\
&                         &&                        4 &&                           4 &&           370  &    0.8$\times$ &&               10.8 &   3.3$\times$ &&                 0.174  &  2.3$\times$ \\
&                         &&                        8 &&                           8 &&           351  &    0.9$\times$ &&               22.8 &   6.9$\times$ &&                 0.277  &  3.7$\times$ \\
&                         &&                       16 &&                          16 &&           400  &    0.8$\times$ &&               40.0 &  12.2$\times$ &&                 0.356  &  4.8$\times$ \\
&                         &&                       32 &&                          32 &&         8,777  &   0.03$\times$ &&                3.6 &   1.1$\times$ &&                 0.056  &  0.7$\times$ \\
\cmidrule{3-15}
& \multirow{2}{*}{FQ-SD}  &&                        1 &&                           1 &&           377  &    0.8$\times$ &&               2.7  &   0.8$\times$ &&                 0.075  &  1.0$\times$ \\
&                         &&                       16 &&                          16 &&           381  &    0.8$\times$ &&      \textbf{42.0} &  12.8$\times$ &&         \textbf{0.751} & 10.0$\times$ \\
\cmidrule{2-15}
& \multirow{5}{*}{\sqcpu} &&                        2 &&                           1 &&           160  &    1.9$\times$ &&                6.2 &   1.9$\times$ &&                 0.122  &  1.6$\times$ \\
&                         &&                        4 &&                           1 &&            92  &    3.3$\times$ &&               10.9 &   3.3$\times$ &&                 0.176  &  2.4$\times$ \\
&                         &&                        8 &&                           1 &&            63  &    4.8$\times$ &&               15.8 &   4.8$\times$ &&                 0.203  &  2.7$\times$ \\
&                         &&                       16 &&                           1 &&            61  &    5.0$\times$ &&               16.4 &   5.0$\times$ &&                 0.171  &  2.3$\times$ \\
&                         &&                       32 &&                           1 &&            62  &    4.9$\times$ &&               16.1 &   4.9$\times$ &&                 0.166  &  2.2$\times$ \\
\cmidrule{3-15}
& \multirow{2}{*}{FD-SQ}  &&                        1 &&                           1 &&           327  &    0.9$\times$ &&                3.1  &  0.9$\times$ &&                 0.093  &  1.2$\times$ \\
&                         &&                       16 &&                           1 &&    \bfseries{21} &   14.5$\times$ &&       \bfseries{47.9} & 14.5$\times$ &&         \bfseries{0.815} & 10.9$\times$ \\

\midrule
\midrule
& \seq                    &&                        1 &&                           1 &&         1,326  &             -- &&                 0.8 &             -- &&               0.017  &           -- \\
\cmidrule{3-15}
\parbox[t]{2mm}{\multirow{14}{*}{\rotatebox[origin=c]{90}{\yfcc}}}
& \multirow{5}{*}{\bcpu}  &&                        2 &&                           2 &&         1,570  &    0.8$\times$ &&                 1.3 &    1.7$\times$ &&               0.025  &  1.5$\times$ \\
&                         &&                        4 &&                           4 &&         1,836  &    0.7$\times$ &&                 2.2 &    2.9$\times$ &&               0.036  &  2.1$\times$ \\
&                         &&                        8 &&                           8 &&         2,057  &    0.6$\times$ &&                 3.9 &    5.2$\times$ &&               0.048  &  2.8$\times$ \\
&                         &&                       16 &&                          16 &&         2,263  &    0.6$\times$ &&                 7.1 &    9.4$\times$ &&               0.064  &  3.7$\times$ \\
&                         &&                       32 &&                          32 &&        36,261  &   0.04$\times$ &&                 0.9 &    1.2$\times$ &&               0.010  &  0.6$\times$ \\
\cmidrule{3-15}
& \multirow{2}{*}{FQ-SD}  &&                        1 &&                           1 && \textbf{1,494} &    0.9$\times$ &&                0.7  &    0.9$\times$ &&               0.019  &  1.1$\times$ \\
&                         &&                       16 &&                          16 &&         1,545  &    0.9$\times$ &&       \textbf{10.4} &   13.7$\times$ &&       \textbf{0.199} & 11.5$\times$ \\
\cmidrule{2-15}
& \multirow{5}{*}{\sqcpu} &&                        2 &&                           1 &&            736 &    1.8$\times$ &&                 1.4 &    1.8$\times$ &&               0.027  &  1.6$\times$ \\
&                         &&                        4 &&                           1 &&            423 &    3.1$\times$ &&                 2.4 &    3.1$\times$ &&               0.039  &  2.3$\times$ \\
&                         &&                        8 &&                           1 &&            285 &    4.6$\times$ &&                 3.5 &    4.6$\times$ &&               0.047  &  2.7$\times$ \\
&                         &&                       16 &&                           1 &&            260 &    5.1$\times$ &&                 3.8 &    5.1$\times$ &&               0.041  &  2.4$\times$ \\
&                         &&                       32 &&                           1 &&            260 &    5.1$\times$ &&                 3.9 &    5.1$\times$ &&               0.039  &  2.3$\times$ \\
\cmidrule{3-15}
& \multirow{2}{*}{FD-SQ}  &&                        1 &&                           1 &&         1,348  &    1.0$\times$ &&                0.7  &    1.0$\times$ &&               0.023  &  1.3$\times$ \\
&                         &&                       16 &&                           1 &&    \textbf{85} &   15.5$\times$ &&       \textbf{11.7} &   15.5$\times$ &&       \textbf{0.201} & 11.6$\times$ \\

\midrule
\midrule
& \seq                    &&                        1 &&                           1 &&           274  &            -- &&                 3.7  &            -- &&                0.083  &           -- \\
\cmidrule{3-15}
\parbox[t]{2mm}{\multirow{14}{*}{\rotatebox[origin=c]{90}{\treccast}}}
& \multirow{5}{*}{\bcpu}  &&                        2 &&                           2 &&           313  &   0.9$\times$ &&                 6.4  &    1.7$\times$ &&               0.128  &  1.5$\times$ \\
&                         &&                        4 &&                           4 &&           257  &   0.8$\times$ &&                11.2  &    3.1$\times$ &&               0.184  &  2.1$\times$ \\
&                         &&                        8 &&                           8 &&           412  &   0.7$\times$ &&                19.4  &    5.3$\times$ &&               0.242  &  2.8$\times$ \\
&                         &&                       16 &&                          16 &&           396  &   0.7$\times$ &&                40.4  &   11.1$\times$ &&               0.367  &  3.7$\times$ \\
&                         &&                       32 &&                          32 &&        30,049  &  0.01$\times$ &&                 1.1  &    0.3$\times$ &&               0.012  &  0.6$\times$ \\
\cmidrule{3-15}
& \multirow{2}{*}{FQ-SD}  &&                        1 &&                           1 &&           290  &   0.9$\times$ &&                 3.4  &    0.9$\times$ &&               0.098  &  1.2$\times$ \\
&                         &&                       16 &&                          16 &&   \textbf{289} &   0.9$\times$ &&        \textbf{55.3} &   15.1$\times$ &&       \textbf{0.927} & 11.2$\times$ \\
\cmidrule{2-15}
& \multirow{5}{*}{\sqcpu} &&                        2 &&                           1 &&           157  &   1.7$\times$ &&                 6.4  &    1.7$\times$ &&               0.126  &  1.6$\times$ \\
&                         &&                        4 &&                           1 &&            81  &   3.4$\times$ &&                12.3  &    3.4$\times$ &&               0.200  &  2.3$\times$ \\
&                         &&                        8 &&                           1 &&            53  &   5.2$\times$ &&                18.0  &    5.2$\times$ &&               0.244  &  2.7$\times$ \\
&                         &&                       16 &&                           1 &&            50  &   5.5$\times$ &&                20.1  &    5.5$\times$ &&               0.210  &  2.4$\times$ \\
&                         &&                       32 &&                           1 &&            51  &   5.4$\times$ &&                19.7  &    5.4$\times$ &&               0.199  &  2.3$\times$ \\
\cmidrule{3-15}
& \multirow{2}{*}{FD-SQ}  &&                        1 &&                           1 &&           256  &   1.1$\times$ &&                 3.9  &    1.1$\times$ &&               0.119  &  1.4$\times$ \\
&                         &&                       16 &&                           1 &&  \textbf{16.5} &  16.6$\times$ &&        \textbf{60.7} &   16.6$\times$ &&       \textbf{0.988} & 11.9$\times$ \\
\bottomrule
\end{tabular}
\end{table*}

\paragraph*{Computational efficiency (RQ1)}
Table~\ref{table:results-rq1} reports the results of our comparisons when retrieving a large number of nearest neighbors. The analysis presented refers to a scenario where $k = 1,024$ for all the cases tested except for the FQ-SD solution with $16$ workers, where $k = 64$ due to the considerations discussed in Section~\ref{sec:logicalconf}. 
The experimental results show \bcpu increases throughput up to $40$, $7.1$, and $40.4$ queries per second on \gist, \yfcc and \treccast, respectively, with a consequent scale-up factor of $12.2\times$, $9.4\times$ and $11.1\times$ w.r.t. the \seq baseline. On the other side, \bcpu does not improve the latency figures w.r.t. the \seq baseline, where we observe a small degradation of the query response time. This is expected as we are employing multi-threading parallelism that introduces the possibility of significantly increasing throughput with a small overhead required to handle threads paid on a per-query level. On the other side, \sqcpu significantly reduces latency of up to $5.0\times$, $5.1\times$ and $5.5\times$ w.r.t. \seq on \gist, \yfcc and \treccast, respectively, while also increasing the overall query throughput. 
Note that our two \knn multi-threaded solutions on CPU show a significant degradation of the performance when increasing the number of workers to more than $16$. Again, this is expected, as we are running experiments on an Intel CPU with $16$ physical cores. By increasing the number of threads to more than the actual number of physical cores, the overhead of handling threads becomes the dominant cost observed in query latency on both datasets. Regarding our FPGA architectures, i.e., FQ-SD and FD-SQ, the experimental results show that they always outperform the CPU competitors on all metrics. 
Specifically, our FQ-SD solution achieves the best throughput in terms of queries per second (up to $42.0$, $10.4$, and $55.3$ on \gist, \yfcc and \treccast, respectively) while achieving in this scenario comparable figures in terms of latency w.r.t. the best-performing method. FD-SQ achieves the same result that allows achieving the best-observed latency over competitors (with scale-up factors of up to $14.5\times$, $15.5\times$ and $16.6\times$ on \gist, \yfcc and \treccast) with also the best throughput (up to $16.6\times$). 
Thus, we can conclude that our proposed architectures can significantly outperform sequential solutions on CPU and, more interestingly, parallel CPU-based implementations by an interesting margin on all three experimental scenarios regarding computational efficiency.

\paragraph*{Energy efficiency (RQ2)}
Table~\ref{table:results-rq1} reports, on top of latency and throughput, the energy efficiency analysis achieved by comparing our FQ-SD and FD-SQ solutions against \seq, \bcpu, and \sqcpu in terms of queries per Joule spent (\# queries/Joule). The experimental settings and scenarios are similar to what was reported in RQ1. On top of increased computational efficiency (RQ1), both our proposed architectures significantly outperform the competitors regarding energy efficiency. Specifically, FQ-SD more than doubles the competitors in terms of energy efficiency and saves energy up to $11.2\times$ w.r.t. \sqcpu. Similar considerations can be made for the FD-SQ solution, achieving up to $11.9\times$ energy saving w.r.t. \sqcpu and more than $4\times$ w.r.t. the \sqcpu natural competitor.
The energy efficiency figures show that FPGA-based solutions for \knn are extremely efficient regarding the number of queries per Joule spent. Our proposed architectures can significantly outperform sequential and parallel CPU-based solutions.

\begin{table*}
\caption{Experimental results on the \cast dataset by varying $k$. We also report the scale-up factor achieved by each method w.r.t. \seq for latency, throughput, and energy efficiency.} \label{table:results-rq2}
\centering
\begin{tabular}{clrrrrrrrrr}
\toprule
\multirow{2}{*}{Dataset}
& \multirow{2}{*}{Method} & \multicolumn{1}{c}{\multirow{2}{*}{K}} & \multicolumn{1}{c}{\multirow{2}{*}{Workers}} & \multicolumn{1}{c}{Batch}   &    \multicolumn{2}{c}{Latency}  &  \multicolumn{2}{c}{Throughput}    & \multicolumn{2}{c}{Energy Efficiency} \\
&                         &                                        &                                              & \multicolumn{1}{c}{Size}    &    \multicolumn{2}{c}{(msec/query)}   &  \multicolumn{2}{c}{(Queries/sec)} & \multicolumn{2}{c}{(Queries/J)}       \\
\toprule
\parbox[t]{2mm}{\multirow{13}{*}{\rotatebox[origin=c]{90}{\treccast}}}
& \multirow{2}{*}{\seq}   & 1024 &  1 &  1 &           304 &           -- &          3.3  &           -- &          0.075 &           -- \\
&                         &    1 &  1 &  1 &           255 &  1.2$\times$ &          3.9  &  1.2$\times$ &          0.090 &  1.2$\times$ \\
\cmidrule{2-11}
& \multirow{4}{*}{FQ-SD}  &   64 & 16 & 16 &          289  &  1.1$\times$ &         55.3  & 16.8$\times$ &         0.927  & 12.4$\times$ \\
&                         &   22 & 19 & 19 &          311  &  1.0$\times$ &         61.1  & 18.5$\times$ &         1.029  & 13.7$\times$ \\
&                         &   10 & 22 & 22 &          286  &  1.0$\times$ &         76.9  & 23.3$\times$ &         1.158  & 15.4$\times$ \\
&                         &    3 & 24 & 24 &          298  &  1.0$\times$ &         80.5  & 24.4$\times$ &         1.135  & 15.1$\times$ \\
\cmidrule{2-11}

& \multirow{4}{*}{FD-SQ}  & 1024 & 16 &  1 &         16.5  & 18.4$\times$ &         60.7  & 18.4$\times$ &         0.988  & 13.2$\times$ \\
&                         &  418 & 19 &  1 &         13.5  & 22.5$\times$ &         74.2  & 22.5$\times$ &         1.170  & 15.6$\times$ \\
&                         &  200 & 22 &  1 &         12.0  & 25.3$\times$ &         83.5  & 25.3$\times$ & \textbf{1.266} & 16.9$\times$ \\
&                         &   72 & 24 &  1 & \textbf{11.5} & 26.4$\times$ & \textbf{86.8} & 26.3$\times$ &         1.253  & 16.7$\times$ \\
\cmidrule{2-11}

& GPU & 72 & 6,912 &  1 & 93.7 & 3.2$\times$ & 10.6 & 3.2$\times$ & 0.028 & 2.6$\times$ \\

\bottomrule
\end{tabular}
\end{table*}

\paragraph*{Trading-off degree of parallelism and cutoff $k$ (RQ3)}
Table \ref{table:results-rq2} reports the experimental results on the \treccast dataset when retrieving a variable number of nearest neighbors $k$. Given the limited amount of resources available on the FPGA programmable logic, we can further improve the performance by trading off the degree of parallelism with the cutoff $k$ of the \knn search. In particular, in many different use cases, e.g., conversational search, it is not required to retrieve a high number of results. We can thus exploit a lower cutoff $k$ to increase the degree of parallelism available for the computation. We compare FQ-SD and FD-SQ against \sqcpu by reporting the latency, throughput, and energy efficiency of the solutions tested. 
The experimental results confirm this intuition on both the FPGA architectures. Indeed, by lowering the cutoff $k$ from $64$ to $3$, we can increase the parallelism of the FQ-SD solution from $16$ to $24$, resulting in an improvement of the throughput from $55.3$ to $80.5$ queries/sec ($+45\%$). The latency of the several combinations of $k$ and degree of parallelism is almost constant. At the same time, we can observe a consistent save in terms of energy spent per query (up to +25\% of queries per Joule). Similarly, by lowering the cutoff $k$ from $1024$ to $72$, we can increase the parallelism of the FD-SQ solution from $16$ to $24$, thus improving the throughput with a solid $+43\%$ (from $60.7$ to $86.8$). Remarkably, with this architecture, we also observe a decrease in the latency (up to $-30\%$) and a power efficiency improved by $+28\%$. If we consider the \sqcpu solution instead, we note that lowering the cutoff $k$ from $1024$ to $1$ has a reduced impact on the performance, achieving $+10\%$ in throughput, $-16\%$ in latency and $+20\%$ in queries per Joule. Finally, the FPGA architectures, compared to the \sqcpu solution, provide a scale-up factor of up to $26.3\times$ and an energy saving of up to $16.9\times$.

\paragraph*{Comparison with GPU-based \knn}
This experiment compares the best performance achieved with our FPGA-based solutions (FD-SQ, $24$ workers) against a GPU-based \knn retrieval operating in the same scenario. We test our methods against the GPU-based flat index\footnote{The flat index in FAISS is the one that allows us to perform an exact \knn search over a collection of vectors.} available in the FAISS library. We load the flat index in the GPU memory, and we perform an exact \knn search on top of it. Loading the flat index on the GPU requires $28$ GB of GPU memory. The energy needed by the search process is monitored using the \texttt{nvidia-smi} toolkit.

Table~\ref{table:results-rq2} reports the results of the experiments. When tested in the same application scenario, i.e., small batch sizes, the results confirm that our FPGA-based solution achieves lower latencies (8.1$\times$) and higher throughput (8.2$\times$) than the GPU-based implementation. This is expected as GPU architectures are designed for batch, high-throughput applications where the performance is heavily dependent on the available memory and the amount of data, i.e., the queries to be batched on the GPU. The higher the memory and the amount of data available, the better the performance achieved in latency and throughput since modern GPUs have a significant amount of parallel cores performing the computation. However, as the results show, modern GPUs are not the right choice in application scenarios where we have a significant inter-arrival time  and the queries cannot be grouped in large batches. To confirm this intuition, we also tested a scenario where the full set of queries of \treccast, i.e., $6$,$980$ is sent to the GPU in a single batch to perform exact \knn retrieval. In this setting, the $6$,$912$ cores of the GPU achieve a per-query latency of $0$.$806$ msec with a throughput of $1$.$115$ queries per second with an energy efficiency of about $3$ queries per Joule. These figures confirm our intuition: when data and memory are available in a huge quantity, GPUs for \knn are the correct choice due to the high throughput achieved. This is not true in low-latency domains where queries cannot be batched. It is also worth highlighting that the performance provided by the GPU comes at a cost. At the time of the writing of this manuscript, an A100 equipped with 40GB of RAM costs approximately 18,000 Euro while an Alveo U55C FPGA with 16GB of RAM costs approximately one-fourth of that amount.





\paragraph*{Comparison with CHIP-KNN}
Given the difficulties in compiling and executing the source code of CHIP-KNN on our FPGA board, in this section, we present a comparative analysis of the throughput (GBytes/sec) reported by Lu~\emph{et al.}~\cite{9415564}\footnote{The authors of such a work refer to throughput as \emph{bandwidth}.}. We compare it with the throughput of our best-performing FPGA solution, optimizing it (FD-SQ). Before proceeding with the comparison, it is important to highlight two key aspects emerging from the experimental evaluation proposed by Lu~\emph{et al.}~\cite{9415564}: i) the dimensionality of the feature vectors; ii) the cutoff $k$. Regarding the first aspect, Lu~\emph{et al.} evaluate their solution by limiting the feature vector dimension $D$ in the range between $2$ and $128$, thus restricting the analysis to vectors of smaller sizes than the ones used in our experimentation. The smallest vectors used in our experimental analysis are the ones of the \treccast dataset having $769$ dimensions each. Moreover, the throughput reported is strongly dependent on $D$. Indeed the best performance reported by Lu~\emph{et al.} is achieved with a small feature vector of size $16$, and it rapidly falls to $115$ GB/s when employing vectors of size $128$. The throughput of CHIP-KNN is thus expected to be even lower with a vector of size $D \geq 769$ like the ones employed in our use cases. On the contrary, the throughput of our best-performing FPGA solution is almost independent of $D$. Indeed, it is $192$ GB/s on \treccast, exploiting vectors of size $769$, and it keeps almost constant, i.e., $187$ GB/s, on \yfcc with vectors of size $4$,$096$. Regarding the second aspect, i.e., the cutoff $k$, the experiments by Lu~\emph{et al.} analyze only small cutoffs, ranging between $5$ and $20$. On the contrary, our analysis is performed with a large cutoff value, i.e., $k = 1024$. Moreover, as already shown in RQ2, by lowering the cutoff $k$ we can increase the resulting parallelism of our solutions, thus achieving higher throughput figures. In detail, with $k = 1024$, the FD-SQ architecture achieves a bandwidth of $190$ GB/s, while with $k = 72$ the throughput of the solution increases up to $257$ GB/s. To sum up, although a direct comparison between our solutions and CHIP-KNN is not possible due to the impossibility of compiling and running the original code provided by Lu~\emph{et al.} on our FPGA board, we can safely claim that our solutions: i) provides higher throughput numbers, ii) are extremely optimized towards throughput or latency, and iii) successfully deal with two different exact \knn application scenarios, i.e., streaming vs. in-memory.

\vspace{-2mm}
\section{Conclusions and Future Work}
\label{sec:conclusions}

This research investigates the benefits of using Field Programmable Gate Array (FPGA) devices for energy-efficient exact \knn search in high-dimensional spaces. To this end, we designed and evaluated two architectures for \knn search on FPGA: FQ-SD and FD-SQ. While the former architecture optimizes throughput by processing in parallel the queries of a batch over a streamed dataset not fitting into the FPGA memory, the latter is designed to reduce the search latency by processing each \knn incoming query in parallel over an in-memory dataset, which is crucial for client-side applications where low latency is a key requirement.
%
%
We conducted experiments on three publicly available datasets for image and text retrieval having feature vectors ranging from $769$ to $4$,$960$ dimensions. We evaluated the efficiency of our solutions against state-of-the-art CPU and GPU \knn competitors in terms of throughput, latency, and energy consumption. 
The results of the experimental assessment show that  our FPGA architectures achieve the best throughput in terms of queries per second---up to $55.3$ on \treccast---or the best-observed latency over CPU-based competitors with scale-up factors of up to $16.6\times$ on the same dataset. Similar considerations can be made regarding energy efficiency, where we showed that we achieve up to $11.9\times$ energy saving w.r.t. \sqcpu. Moreover, we also showed that FD-SQ outperforms CHIP-KNN, the state-of-the-art FPGA \knn solution. Finally, concerning the state-of-the-art FAISS \knn library for GPUs, we observed that our FPGA solution remarkably outperforms even a high-end GPU in application scenarios where we cannot batch the queries in large sets to exploit the power of all GPU cores fully. 

\paragraph*{Future Work}
In future work, we plan to investigate the impact of quantization techniques on the efficiency and effectiveness of our FPGA solutions. Quantization can be applied to the vectors to reduce the memory required to store them. Consequently, quantization techniques can improve the overall throughput of the solutions in terms of queries per second while degrading the overall effectiveness of the \knn search due to the approximation introduced on the representation. The evaluation of this effectiveness/latency/energy-efficiency trade-off is an interesting research line we intend to investigate.
We also plan to investigate the scalability of our approach when multiple FPGAs within a single system are employed. 

\section*{Acknowledgments}
This research has been partially funded by the European Union’s Horizon Europe research and innovation program EFRA (Grant Agreement Number 101093026). Views and opinions expressed are however those of the authors only and do not necessarily reflect those of the European Union or European Commission-EU. Neither the European Union nor the granting authority can be held responsible for them.

\bibliographystyle{ACM-Reference-Format}
\bibliography{biblio}

\end{document}